\documentclass[pra,twocolumn,reprint,groupedaddress,showpacs]{revtex4-1}
\usepackage[utf8]{inputenc} 
\usepackage{graphicx}
\usepackage{amsmath}
\usepackage{units}

\usepackage{color}

\begin{document}
\title{Observation of low-field Fano-Feshbach resonances in ultracold gases of dysprosium}
\author{Kristian Baumann}
\author{Nathaniel Q. Burdick}
\author{Mingwu Lu}
\author{Benjamin L. Lev}
%\email{BenLev@Stanford.edu}
\affiliation{Department of Applied Physics, Stanford University, Stanford CA 94305}
\affiliation{Department of Physics, Stanford University, Stanford CA 94305}
\affiliation{E. L. Ginzton Laboratory, Stanford University, Stanford CA 94305}

\date{\today}

\begin{abstract}
We report the observation of resonance-like loss in the trap population of ultracold dysprosium as a function of magnetic field, which we attribute to anisotropy-induced Fano-Feshbach resonances arising from Dy's large magnetic dipole moment and nonzero electronic orbital angular momentum.  We  recorded these resonances for four different isotopes, three bosonic and one fermionic, over a field range of \unit[0-6] G and show that the number of resonances changes significantly as a function of temperature, even in the nK regime.  Most of the observed resonances are of very narrow width. The fermionic isotope, unlike its bosonic counterparts, possesses nonzero nuclear spin and exhibits a much higher density of resonances.
\end{abstract}
 \maketitle

%Introduction
Fano-Feshbach resonances have become an essential tool for tuning the scattering properties between atoms and molecules in the single--to--few partial wave regime, typically achieved at very low temperatures~\cite{Chin2010,Inouye1998}.  Tuning a magnetic field near such resonances  has proven to be an important tool  for achieving efficient evaporative cooling, especially  for creating  large Bose-Einstein condensates (BECs) of atoms whose scattering lengths at zero field are unfavorable for providing rethermalization with low trap loss~\cite{Cornish2000,Weber2003}. Moreover, these resonances provide experimental access to intriguing three-body Efimov states \cite{Kraemer2006}, loosely-bound molecules~\cite{Kohler2006}, and strongly-correlated many-body states of matter, such as found at the BEC-BCS crossover~\cite{Giorgini2008,Bloch2008}. 

The scattering properties of highly magnetic lanthanide atoms are subject to both the electrostatic anisotropy of the nonzero orbital angular momentum ground state and the anisotropy of the dipolar interaction, and as a consequence, they are predicted to exhibit a much higher density of resonances than in systems of alkali atoms~\cite{Petrov2012}. Indeed, Fano-Feshbach resonances in highly magnetic atoms have been observed in chromium ($\mu=6$ Bohr magnetons, $\mu_B$)~\cite{Werner2005} and very recently in erbium (7 $\mu_{B}$)~\cite{Aikawa2012,Frisch2013}. We report the observation of low-field Fano-Feshbach resonance spectra for   ultracold dysprosium (10 $\mu_{B}$):  We investigate the influence of temperature, which in contrast to alkali atoms, has significant effects  well below \unit[1]{$\mu$K}, and we present data for three bosonic isotopes ($^{160}$Dy, $^{162}$Dy, and $^{164}$Dy) and one fermionic isotope ($^{161}$Dy). Overall, the density of resonances per gauss of magnetic field largely exceeds the density
expected by theory~\cite{Petrov2012} based on modeling in the zero temperature limit, and we show that the density is even larger for the fermionic isotope, as also recently observed in fermionic Er~\cite{Frisch2013}.

% Theoretical background - Svetlanas Paper
The collisional properties of lanthanide atoms at ultralow temperatures are fundamentally different than alkali atoms'.  Dysprosium has an open $f$-shell submerged beneath a closed outer $s$-shell, and bosonic Dy collides under a web of complex molecular potentials involving as many as 153 Born-Oppenheimer surfaces~\cite{Kotochigova2011} due to the large spin of its ground state.  Bosonic Dy possesses a total electronic angular momentum of $J=S+L=8$, with an orbital angular momentum $L=6$ (electronic spin $S$), leading to a large electrostatic anisotropy~\cite{Connolly:2010jb} that greatly affects scattering properties.  Moreover, the long-range and anisotropic dipole-dipole interaction arising from this large spin is larger than any other fermionic atom---and equal only to bosonic terbium's~\cite{Martin:1978}.  This interaction, along with the large electrostatic anisotropy, couples incoming partial waves, inducing ultracold collisional resonances that arise due to rotating bound states~\cite{Petrov2012}.  Needless to say,  full theoretical treatment challenges our computational capabilities, making these experimental data all the more valuable to understanding these unusual dipolar gases which scatter via multiple partial waves no matter how low the temperature. 

Recent numerical calculations have elucidated the fact that the positions and widths of resonances only converge if a large number of partial waves are taken into account, i.e., attributing a single partial-wave scattering channel to a resonance, as is the case with alkali atoms, is no longer possible~\cite{Petrov2012}.  
Intuitively, though as yet unsupported by calculation, the nonzero nuclear spin ($I=5/2$, $F=J+I=21/2$) of fermionic Dy should lead to a higher density of collisional resonances than their $I=0$ bosonic counterparts---due to the expanded manifold of collisional potentials---and we  show that this intuition is indeed supported by the $^{161}$Dy data. These complex atoms  provide a testbed for  exploring ultracold molecular collisions, though without complications arising from  chemical reactions~\footnote{J. Bohn, private communication (2010).}.

%Experimental sequence
The experimental procedure we employ to prepare ultracold samples of Dy has been reported  in detail in Refs.~\cite{lu2011,lu2012}. In short, we load a magneto-optical trap (MOT) from a Zeeman slower, both at \unit[421]{nm}. A subsequent blue-detuned, narrow-line MOT at \unit[741]{nm} provides a cloud of $\sim$$1\times10^8$ atoms at \unit[1-3]{$\mu$K}, and at a peak density of $\sim$\unit[$10^{11}$]{cm$^{-3}$} (the atom number for the fermionic isotope $^{161}$Dy is \unit[$3\times10^7$]{atoms} due to its lower natural abundance). The atomic cloud is subsequently loaded into an optical dipole trap (ODT) composed of a single laser beam with an initial power of \unit[5]{W}, a wavelength of \unit[1064]{nm}, and a waist of 24  $\times$ 22 $\mu$m$^2$. To enhance the loading efficiency, we expand the beam horizontally with an acousto-optical modulator to an aspect ratio of $\sim$5 \cite{Baier2012}.  Radio-frequency (rf) induced rapid adiabatic passage transfers the atoms into their lowest magnetic sub-level---$(J=8,m_J=-8)$ for the bosons and $(F=21/2,m_F=-21/2)$ for the fermion---resulting in 3-10$\times$$10^6$ atoms at a temperature of \unit[5-10]{$\mu$}K after compression in the ODT, depending on the isotope. The atoms are naturally spin polarized at this stage to better than 95\%,  determined through a Stern-Gerlach measurement and absorption imaging; subsequently, any energetically higher-lying population rapidly decays to the ground state via dipolar relaxation~\cite{Hensler2003}.  We form a crossed-beam ODT by overlapping an additional laser beam of \unit[3]{W} at a wavelength of \unit[1064]{nm} and a waist of 34 $\times$ 34~$\mu$m$^2$ at an intersection angle of $45^\circ$ with respect to the initial ODT. Shifting the focus of this beam away from the focus of the initial beam allows us to shape the trapping frequencies to our needs. We implement forced evaporative cooling by decreasing the intensity of both ODTs within \unit[4]{s}. This evaporative cooling sequence is also efficient for the singly trapped spin-polarized fermionic isotope ($^{161}$Dy) due to dipolar scattering, as reported in Ref.~\cite{lu2012}, and which also has recently been demonstrated for $^{167}$Er~\cite{Aikawa2013}.

% magnetic field calibration
High-resolution atom loss spectroscopy of the Fano-Feshbach resonances requires precise control of the magnetic field at the position of the atoms. To generate those fields, we use a set of coils in near-Helmholtz configuration providing a field of $\sim$\unit[2]{G/A}, where the current is set and stabilized with an analog feedback circuit allowing us to apply a field of any value within the range of \unit[0-6]{G} along gravity. We have calibrated the field and verified its stability over several days by the following rf spectroscopy procedure: Transitions between magnetic sublevels are driven by a weak rf field. Atoms eventually decay towards the magnetic ground state by spin-changing dipolar collisions \cite{Hensler2003}, thereby converting their Zeeman energy into kinetic energy. This energy is larger than the depth of the ODT and the atoms are lost from the trap. We map  the resonant loss feature at fixed magnetic field by changing the frequency of the rf field between consecutive runs of the experiment. A fit to the data determines the position of the resonance to within better than \unit[1]{mG}. For the fermion, we have taken the quadratic Zeeman shift into account by numerically diagonalizing the hyperfine Hamiltonian. Repeating this measurement for different absolute magnetic fields allows us to calibrate the fields and to observe a very slow drift of offset fields of less than \unit[5]{mG} over a week.

% describe the procedure to map out resonances
In order to map  the spectrum of Fano-Feshbach resonances as a function of magnetic field, we quickly ($<$\unit[10]{ms}) ramp the magnetic field to the desired target value after the initial preparation sequence. The power of the ODT beams is then dropped by $\sim$30\%, which leads to an immediate loss of atoms and an out-of-equilibrium velocity distribution of the atomic ensemble. Subsequently, the remaining atoms rethermalize, resulting in cooling and further atom loss.  The center and width of Fano-Feshbach resonances are identified by measuring the temperature and atom number after a fixed holding time. Measurements are performed via absorption imaging with a 10-ms time-of-flight.
In order to ensure comparable conditions among datasets, we have adjusted the loading, cooling, and probing sequence slightly for each individual isotope. We  report the final trap geometry, holding time, atomic density, and temperature for each isotope measurement in the following sections.

\begin{figure}[]
\includegraphics[width=1.\columnwidth]{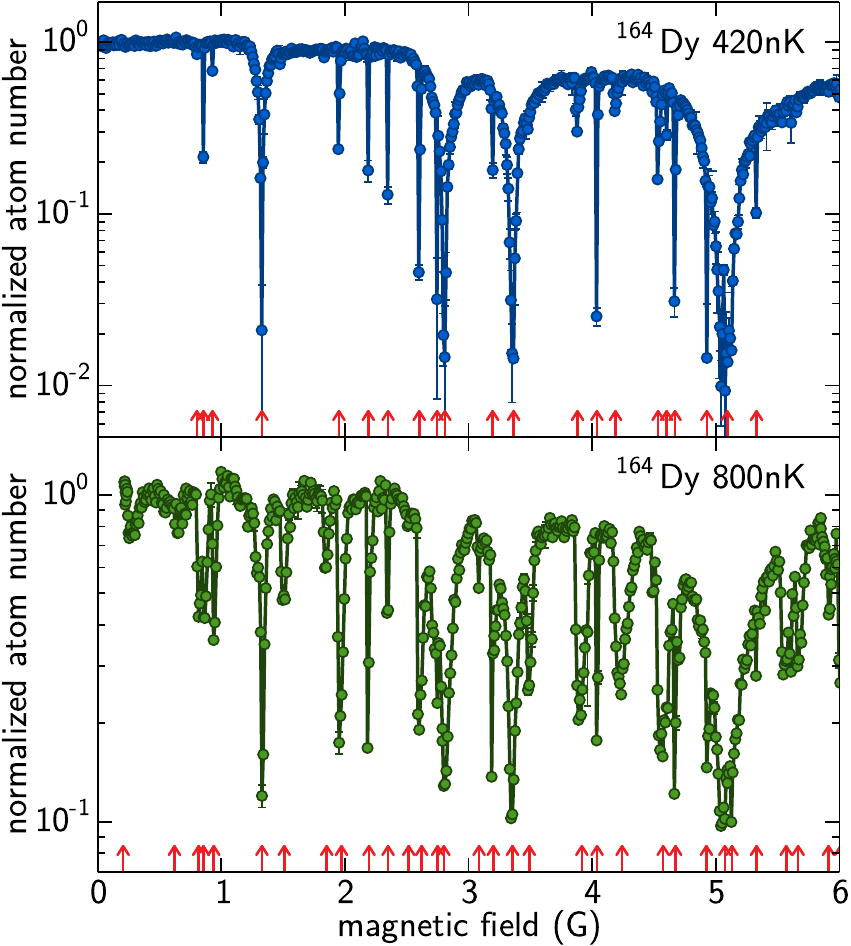}
\caption{(Color online) Atom-loss spectrum for bosonic $^{164}$Dy. The top panel displays data taken with a $\sim$\unit[420]{nK} sample showing a large number of Fano-Feshbach resonances. The lower panel shows a measurement of the same isotope $^{164}$Dy at higher temperature $\sim$\unit[800]{nK}. New resonances appear in the warmer sample.  The arrows show the position of the resonances obtained by fitting a Gaussian function to each individual feature.  Experimental parameters in main text.} \label{fig1}
\end{figure}
% describe the data taking mode + random + reference
We start by presenting data for the bosonic isotope $^{164}$Dy in Fig.~\ref{fig1}. The top panel shows the normalized atom number as a function of magnetic field after the drop in ODT trap depth.   Final trap frequencies are [266, 76, 299]~Hz ($x$-$y$ plane and $\hat{z}$, respectively, where gravity points along $\hat{z}$) and the trap is held for \unit[375]{ms}. For every field value, we repeat the measurement twice and show the mean of those experimental runs with the error bars given by the standard deviation.  All datasets were taken in blocks of \unit[250]{mG}, starting at the lowest fields and in \unit[10]{mG} steps. The 50 runs taken within each block are recorded in random order to counter systematic drifts.  We also repeatedly measure the atom number at a reference magnetic field, which is the same for all blocks, to normalize the atom number.  For the data presented in Fig.~\ref{fig1} (top panel), we use \unit[0.485]{G} as a reference field, where we measure \unit[3$\times10^5$]{atoms} at an average density of \unit[$9\times10^{13}$]{cm$^{-3}$} and a temperature of \unit[420]{nK}.

%describe what we see for cold $^{164}$Dy
The data Fig.~\ref{fig1} (top panel) show a large number of pronounced dips in the atom number as a function of magnetic field, far beyond what is seen in collisions of alkali atoms. We observe 21 Fano-Feshbach resonances below \unit[6]{G} corresponding to an average of more than 3 resonances per gauss. The arrows in the figure show the position of the resonances obtained by fitting a Gaussian function to each individual resonance. Many features show a very small width, $\sim$10 mG. We have repeated the measurement with $^{164}$Dy at a higher temperature, shown in the lower panel of Fig.~\ref{fig1}. The final trap frequencies of the ODT are [291, 85, 328]~Hz, and the ODT is held for \unit[100]{ms}. At the reference magnetic field of \unit[0.535]{G}, we measure \unit[4.4$\times10^5$]{atoms} at an average density of \unit[$8\times10^{13}$]{cm$^{-3}$} and a temperature of \unit[800]{nK}. The spectrum is qualitatively similar to the colder gas;  however, all resonances  are  broadened and new resonance features are apparent. The total number of resonances observed is 32, resulting in a density of $\sim$5 resonances per gauss. 

The broadening with increased temperature suggests  contributions of higher-order partial waves to the collisional process. Indeed, theoretical calculations require the contributions of many partial waves to achieve numerical convergence due to the anisotropy of the dipolar interaction~\cite{Petrov2012,Bohn2009}. The sample temperature affects  the particular contribution weight, resulting in a narrowing and vanishing of resonances. This is in contrast to alkali atoms in which the typical $p$-wave barrier is in the mid--to--high $\mu$K regime, below which higher partial-wave resonances cannot be observed and the spectrum  no longer changes as a function of temperature~\cite{Chin2010}.
The width of the observed resonances is not purely determined by the intrinsic width of the corresponding Fano-Feshbach resonance but is likely broadened by the use of a harmonically trapped gas and the associated spatially dependent atom loss, in addition to the possible presence of a small field gradient.

% Why do not we see the temperature peak
For alkali atoms, the  van der Waals interaction-dominated scattering cross section shows a zero-crossing associated with a vanishing scattering length near each Fano-Feshbach resonance. The atomic gas is effectively non-interacting at thermal temperatures, thus inhibiting  rethermalization after a sudden, non-adiabatic change in the trapping conditions. This effect has been observed with alkali atoms \cite{Zaccanti2006} and also with Er \cite{Aikawa2012}.  The temperature measured in the more strongly dipolar Dy gases employed here, by contrast, may differ from the standard picture of a peak arising in the temperature vs magnetic field at each zero-crossing.  This may arise from the non-vanishing contribution to the overall elastic scattering from dipolar scattering, which for Dy is expected to provide a large magnetic field-independent cross section of \unit[$2.4\times10^{-12}$]{cm$^{-2}$} \cite{Bohn2009}, which scales as $\mu^4$.  Indeed, we do not observe this temperature feature and were not able to observe a considerable change in the thermalization rates of the atomic ensemble near Fano-Feshbach resonances.  Though further investigation is required, we argue that the scattering properties of Dy are generally dominated by dipolar scattering and minimizing the van der Waals contribution insignificantly alters rethermalization. The thermalization rates could be enhanced by greatly increasing the van der Waals contribution. However, this regime, close to a Fano-Feshbach resonance, is also accompanied by increased inelastic three-body collisions, and the concomitant trap loss likely masks the altered thermalization rate.

\begin{figure}[]
\includegraphics[width=1.\columnwidth]{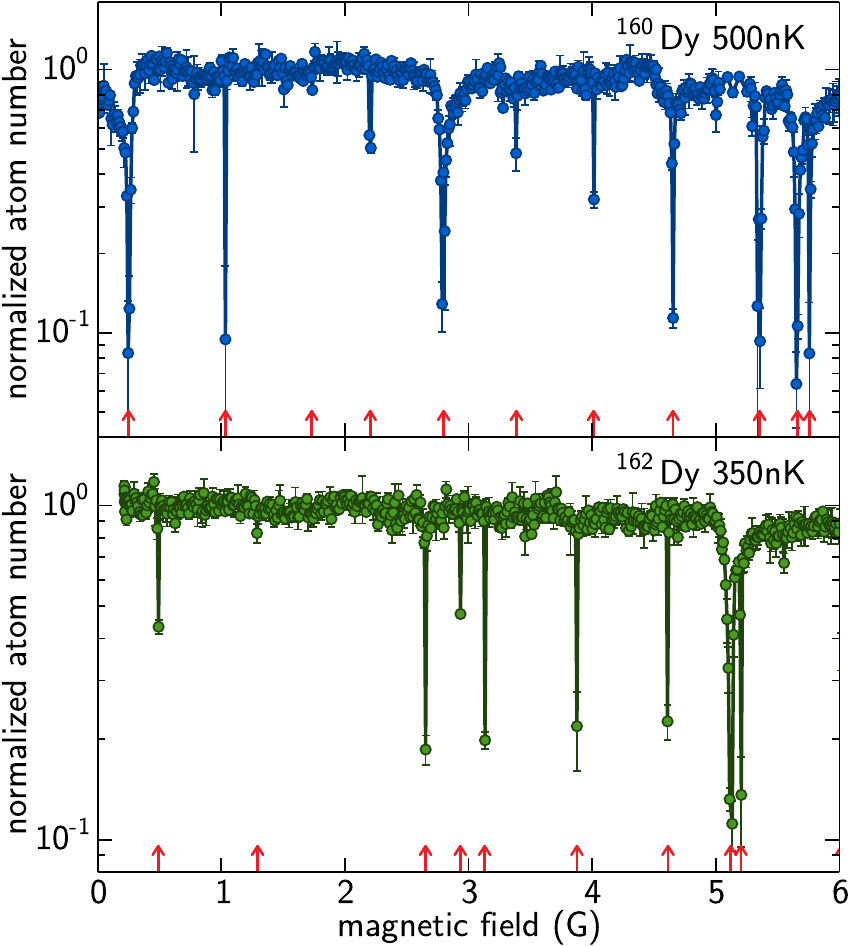}
\caption{(Color online) Atom loss spectra as a function of magnetic field for the bosonic isotopes $^{160}$Dy (top) and $^{162}$Dy (bottom). The spectra lack the broad feature observed in $^{164}$Dy (Fig.~\ref{fig1}), but are otherwise qualitatively similar. The arrows show the position of the resonances obtained by fitting a Gaussian function to each individual feature. Experimental parameters in main text.} \label{fig2}
\end{figure}

% Observation of $^{160}$Dy and 162
In general, one expects the spectrum of Fano-Feshbach resonances to be highly dependent on isotope \cite{Petrov2012}.  The small, few GHz isotope shifts in the energy levels of Dy allows us to easily tune our cooling lasers  between isotopes.  We have repeated the above measurements with two other bosonic isotopes, $^{160}$Dy and $^{162}$Dy, at very low but non-condensed temperatures.  The scanned magnetic field range, resolution, number of experimental runs, and randomization was chosen to be identical to the previous measurement.  Figure~\ref{fig2} (top panel) shows the resulting spectrum for the bosonic isotope $^{160}$Dy: trap frequencies [259, 53, 287]~Hz held for \unit[300]{ms} with \unit[4.9$\times10^4$]{atoms}, \unit[$9.3\times10^{11}$]{cm$^{-3}$} density, and a temperature of \unit[500]{nK}.  Likewise, the lower panel of Fig.~\ref{fig2} plots the $^{162}$Dy data: trap frequencies [251, 7, 276]~Hz held for \unit[375]{ms} with \unit[1.4$\times10^5$]{atoms}, \unit[$1\times10^{13}$]{cm$^{-3}$} density, and a temperature of \unit[350]{nK}.  Both spectra are qualitatively very similar to $^{164}$Dy (Fig.~\ref{fig1}) and show 11 (9) resonances for $^{160}$Dy ($^{162}$Dy), though neither exhibits the relatively wide resonance of  $^{164}$Dy within the 0-6 G scan range.  The density of resonances per gauss  is smaller than the observed Fano-Feshbach density in $^{164}$Dy.  Further investigation with wider field ranges will be necessary to draw  statistically significant conclusions regarding isotope-dependent differences in density.  We note that the relatively small density observed here can be advantageous from a practical point of view, as it allows magnetic field sweeps without crossing resonances and the associated inelastic collisions.  Broad features, as observed in the low-field $^{164}$Dy spectrum, on the other hand, allow the interaction between atoms to be tuned without the stringent field-control requirements associated with 10-mG features.

\begin{figure}
\includegraphics[width=1.\columnwidth]{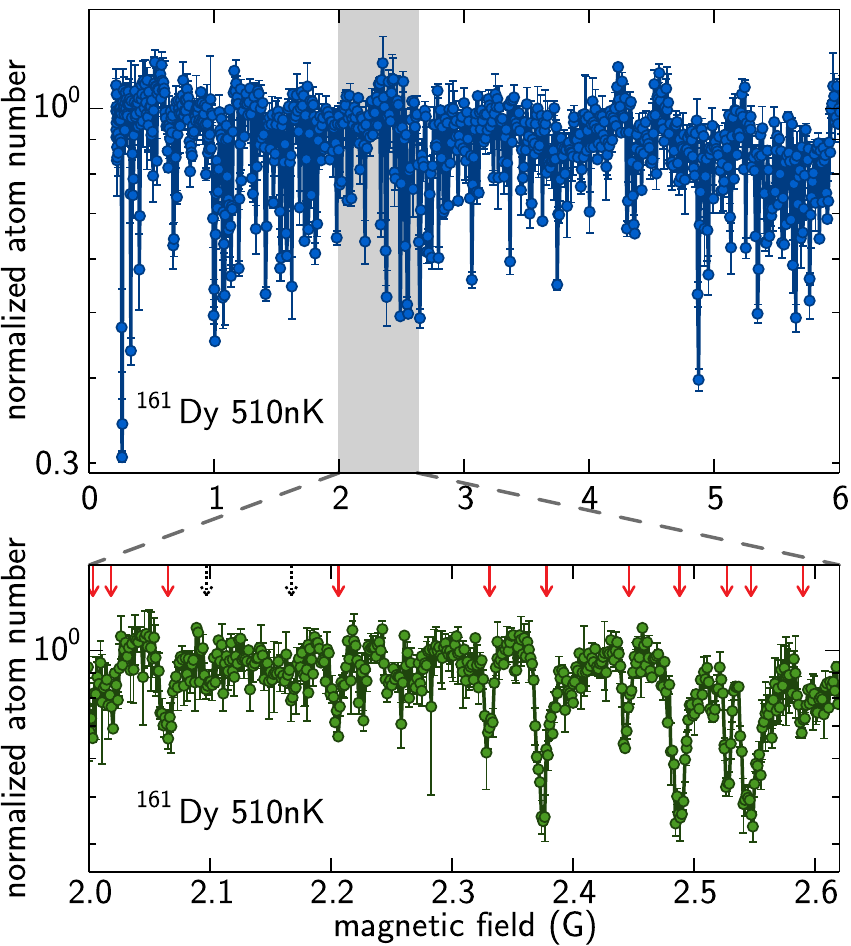}
\caption{(Color online) Atomic loss spectrum for the fermionic isotope $^{161}$Dy. The top panel shows a full scan over \unit[6]{G} whereas the lower panel shows a close-up with much higher resolution.  As apparent from the close-up plot, the ultra-narrow features in the top panel are scattering features. The arrows in the lower panel show the positions of the resonances deduced from the upper panel; the arrows in black (dashed) mark false resonances, i.e., statistical fluctuations in the upper panel not reproduced in the lower panel with high-resolution data. Experimental parameters in main text.} \label{fig3}
\end{figure}

%Observation of $^{161}$Dy
Figure~\ref{fig3} shows data taken with the fermionic isotope $^{161}$Dy. The top panel shows a measurement over the full range of \unit[6]{G} with a resolution of \unit[7.8]{mG}:  trap frequencies are [326, 11, 360]~Hz, ODT held for \unit[240]{ms} with \unit[4.9$\times10^4$]{atoms} at an average density of \unit[$2\times10^{12}$]{cm$^{-3}$} and a temperature of \unit[510]{nK}. The temperature in this configuration is a factor of 5 higher than the Fermi temperature. The spectrum appears much denser than the bosonic counterparts in Figs.~\ref{fig1} and~\ref{fig2}: We observe a very large number of ultra-narrow features and the absence of  broad resonances. The lower panel of Fig.~\ref{fig3} shows a \unit[0.6]{G} sweep with an increased resolution of \unit[1.3]{mG}. The two datasets taken on two consecutive days show excellent agreement and reveal a width to  features that appear only as a single data point in the lower resolution, wide-scan data.  (We have not yet explored the other fermionic isotope $^{163}$Dy since we have found that rapid hyperfine changing collisions induce fast trap loss, requiring the pre-pumping---not yet implemented---of population from the $F=21/2$ state into the ground state $F=11/2$.)

%  comparison between Bosons and Fermions
We observe more than 70 resonances below \unit[6]{G} for  fermionic $^{161}$Dy, yielding a density that exceeds 11 resonances per gauss. This number is much larger than the densities observed for its bosonic counterparts $^{160}$Dy, $^{162}$Dy and $^{164}$Dy, and can qualitatively be explained by the nonzero nuclear spin of the fermionic $^{161}$Dy, which is zero for the bosonic isotopes. The resulting hyperfine coupling provides an additional spin degree of freedom,  greatly increasing the number of available collisional channels.  We calculated the average nearest-neighbor magnetic field separation between resonances in fermionic $^{161}$Dy, but did not observe the Wigner-Dyson distribution of separations, normalized by this mean separation, expected of a system with energy level correlations and whose classical counterpart exhibits ``chaotic''-like sensitivity to Hamiltonian parameter variation~\cite{Mayle2012}.    That the statistics of Fano-Feshbach resonances in this fermionic isotope are more consistent with a Poisson distribution, unlike that recently reported for bosonic Er \cite{Frisch2013}, may simply  be a consequence of the fewer number of resonances recorded, and performing magnetic-field scan ranges beyond 6 G, for both bosonic and fermionic isotopes, would presumably allow one to determine whether resonance in Dy conform to a Wigner-Dyson distribution.

% Conclusion
In conclusion, we have observed Fano-Feshbach resonances for four isotopes of Dy---bosonic and fermionic---in a magnetic field range from 0 to 6 G. These low-field spectra exhibit rich structures,  possessing a density of 2-3 resonances per gauss for the bosonic isotopes and more than 11 resonances per gauss for the fermionic isotope, which exceeds theoretical predictions. The density of resonances changes strongly as a function of temperature, even in the nK regime.  Future work may address the statistics of these resonances.

% Acknowledgments
We thank J. Bohn and S. Kotochigova for enlightening discussions regarding resonance statistics and partial wave contributions. We acknowledge support from the NSF, AFOSR,  and the Quantum Circuits ARO-MURI.

%\bibliography{bibliography.bib}


\end{document}